\documentclass[12pt]{article}
\usepackage{amsmath,amssymb,mathrsfs}
\usepackage{url,hyperref,pict2e}

\title{Boundary Conditions that Remove Certain Ultraviolet Divergences}
\author{
Roderich Tumulka\footnote{Mathematisches Institut,
     Eberhard-Karls-Universit\"at, Auf der Morgenstelle 10, 72076
     T\"ubingen, Germany. E-mail:
     roderich.tumulka@uni-tuebingen.de}
}
\date{March 26, 2021}

\newcommand{\Hilbert}{\mathscr{H}}
\newcommand{\Fock}{\mathscr{F}}
\newcommand{\Q}{\mathcal{Q}}
\newcommand{\sM}{\mathscr{M}}
\newcommand{\sS}{\mathscr{S}}
\renewcommand{\Re}{\mathrm{Re}}
\renewcommand{\Im}{\mathrm{Im}}
\newcommand{\RRR}{\mathbb{R}}
\newcommand{\CCC}{\mathbb{C}}
\newcommand{\PPP}{\mathbb{P}}
\newcommand{\SSS}{\mathbb{S}}
\newcommand{\scp}[2]{\langle #1|#2 \rangle}
\newcommand{\orig}{{\mathrm{orig}}}
\newcommand{\vj}{\boldsymbol{j}}
\newcommand{\vk}{\boldsymbol{k}}
\newcommand{\vn}{\boldsymbol{n}}
\newcommand{\vx}{\boldsymbol{x}}
\newcommand{\vy}{\boldsymbol{y}}

\newcommand{\vB}{\boldsymbol{B}}
\newcommand{\vE}{\boldsymbol{E}}

\newcommand{\valpha}{\boldsymbol{\alpha}}
\newcommand{\vomega}{\boldsymbol{\omega}}
\newcommand{\vzero}{\boldsymbol{0}}
\newcommand{\free}{\mathrm{free}}
\newcommand{\inter}{\mathrm{int}}
\newcommand{\ibc}{\mathrm{IBC}}
\newcommand{\be}{\begin{equation}}
\newcommand{\ee}{\end{equation}}

\newcommand{\un}{\unitlength}
\newcommand{\spacelike}{\begin{minipage}[b]{7\un}
\begin{picture}(6,6)
\put(0,5){\line(1,-1){6}}
\put(0,-1){\line(1,1){6}}
\end{picture}
\end{minipage}}

\begin{document}
\maketitle
\begin{abstract}
In quantum field theory, Hamiltonians contain particle creation and annihilation terms that are usually ultraviolet (UV) divergent. It is well known that these divergences can sometimes be removed by adding counter-terms and taking limits in which an UV cut-off tends to infinity. Here, I review a novel way of removing UV divergences: by imposing a kind of boundary condition on the wave function. These conditions, called interior-boundary conditions (IBCs), relate the values of the wave function at two configurations linked by the creation or annihilation of a particle. They allow for a direct definition of the Hamiltonian without renormalization or limiting procedures. In the last section, I review another boundary condition that serves for determining the probability distribution of detection times and places on a timelike 3-surface.

\medskip

\end{abstract}

\section{Introduction}

Let us take an unusual approach to quantum field theory (QFT): Let us use wave functions. Specifically, let us use a particle-position representation of the quantum state vector. I know there are reservations about such an approach, but let us set aside these reservations for a little while and explore what this may gain us. My message here is that this approach has led to a certain new type of boundary conditions on the wave function, called interior-boundary conditions (IBCs), that make possible UV-finite particle creation and annihilation terms in the Hamiltonian (in some models, at least). 

Is it desirable at all to obtain a Hamiltonian without UV divergence? Many researchers have given up on this aim long ago. Understandably so: If there has been not much progress on the UV divergence problem in a long time, then we will want to move on without solving this problem. But that does not mean that a solution to the problem would not be welcome. Likewise, many physicists have focused on computing scattering matrices, which represent the time evolution from $t=-\infty$ to $t=+\infty$. But that does not mean that the time evolution in between, for finite times, did not exist in nature. For those QFTs for which a well-defined Hamiltonian exists, it seems to me very valuable to know that it does.

In this paper, I explain the idea of IBCs and how it works. I report about some recent progress on providing well-defined Hamiltonians by means of IBCs, mainly concerning non-relativistic models, and about some reasons for thinking that IBCs can also be applied in quantum electrodynamics (QED). It has turned out that Hamiltonians defined with IBCs agree with those obtained through renormalization in the cases where both approaches are available. However, the IBC approach has also been applied successfully to models for which no well-defined Hamiltonian was known (neither through renormalization nor otherwise). In the last section of this paper, I briefly turn to another boundary condition on wave functions relevant to the probability distribution of arrival times or, more precisely, detection times.

In addition to UV divergence, some QFTs also have infrared (IR) divergence problems. But they seem less serious because they usually disappear if 3-space is assumed to have finite volume, and it seems like a real possibility that that is the case in our universe. 

IBCs were first considered by Moshinsky in 1951 \cite{Mosh51a,Mosh51b,Mosh51c}, also for the purpose of modeling particle creation and annihilation but not for dealing with UV divergence. This use was proposed in 2015 \cite{TT15a} and has led to a number of papers on IBCs since.  

In Section~\ref{sec:nonrel}, I discuss IBCs in the non-relativistic case. I include an overview of the literature in Section~\ref{sec:literature}, also covering works on the relativistic case. In Section~\ref{sec:qed}, I explain why there is reason to think that the approach can also be applied to QED, and what the present obstacles are. In Section~\ref{sec:detect}, I describe another boundary condition, an absorbing boundary condition, used for formulating Born's rule for a timelike 3-surface.

\section{Non-Relativistic Case}
\label{sec:nonrel}

So let us consider wave functions in the particle-position representation. In $N$-particle quantum mechanics, the wave function can be regarded as a function $\psi$ on configuration space $\RRR^{3N}$ with values in spin space $S^{\otimes N}$ with $S=\CCC^{2s+1}$ the spin space of a single spin-$s$ particle. For models involving particle creation, $\psi$ should be a function on the configuration space $\Q$ of a variable number of particles,
\be\label{Qdef}
\Q = \bigcup_{n=0}^\infty (\RRR^3)^n\,.
\ee
We could use ordered configurations $(\vx_1,\ldots,\vx_n)$ or unordered configurations $\{\vx_1,\ldots,\vx_n\}$; it will be convenient for us to use ordered ones. On the $n$-particle sector of $\Q$, $\psi$ takes values in $S^{\otimes n}$, so it can be written as $\psi_{s_1...s_n}(\vx_1...\vx_n)$ with $s_j\in\{1,\ldots,2s+1\}$ and $\vx_j\in\RRR^3$. For brevity, we often write $(\vx_1...\vx_n)$ for $(\vx_1,\ldots,\vx_n)$. Furthermore, $\psi$ should be anti-symmetric (for fermions) or symmetric (for bosons) against permutations of the particle positions along with their spin indices. The set of (square-integrable) such wave functions $\psi$ is just the (fermionic or bosonic) Fock space $\Fock$.

\subsection{Simple Model}
\label{sec:model}

Let us consider two particle species, $x$ and $y$, such that $x$-particles (fermions) can emit and absorb $y$-particles (bosons), $x\rightleftarrows x+y$; this is sometimes called the Lee model \cite{Lee54,schweber:1961}. For simplicity, let us focus first on the case of a single $x$-particle fixed at the origin $\vzero\in\RRR^3$; this is sometimes called the van Hove model \cite{vH52,Der03}. Also for simplicity, let us suppose further that all particles, $x$ and $y$, have spin $s=0$ (although in nature all fermions have half-odd spin; spin-statistics theorems do not apply here, as the model is non-relativistic). Then $\psi$ is a complex-valued function of the $y$-configuration, $\psi=\psi(\vy_1...\vy_n)$ with variable particle number $n$, or equivalently a vector in the bosonic Fock space of the $y$-particles. 

The natural formula for a Hamiltonian of free non-relativistic $y$-particles that get created and annihilated at the origin is
\begin{align}\label{Horig}
(H_\orig\psi)^{(n)}(\vy_1...\vy_n)
&= -\frac{\hbar^2}{2m_y} \sum_{k=1}^n \nabla_k^2 \psi^{(n)}(\vy_1...\vy_n) + n E_0\: \psi^{(n)}(\vy_1...\vy_n)\nonumber\\
&\quad + g\, \sqrt{n+1}\: \psi^{(n+1)}(\vy_1...\vy_n,\vzero)\nonumber\\
&\quad + \frac{g}{\sqrt{n}} \sum_{k=1}^n \delta^3(\vy_k) \: \psi^{(n-1)}(\vy_1...\vy_{k-1},\vy_{k+1}...\vy_n)\,,
\end{align}
where $\psi^{(n)}$ means the $n$-particle sector of $\psi$, $m_y>0$ is the mass of a $y$-particle, $E_0\geq 0$ is the energy that needs to be expended for creating a $y$-particle (which would reasonably be $E_0=m_yc^2$, but that is not crucial for what follows), $g\in\RRR$ is a coupling constant governing the strength of the emission and absorption, and $\delta^3$ means the Dirac delta function in 3 dimensions. The first line represents the free time evolution of the $y$s, the second the absorption of $\vy_{n+1}$ by the $x$, and the third the creation of a $y$ by the $x$; the sum over $k$ ensures the permutation symmetry. In different notation, the Hamiltonian can be written as
\be\label{Horig1b}
H_\orig= H_\free+H_\inter
\ee
with the free boson Hamiltonian
\be\label{Horig2a}
H_\free = d\Gamma \Bigl(-\tfrac{\hbar^2}{2m_y}\nabla^2 +E_0\Bigr),
\ee
where $d\Gamma(H_1)$ means the ``second quantization functor'' applied to a 1-particle Hamiltonian $H_1$, and the interaction Hamiltonian
\be\label{Horig2b}
H_\inter = g\: a(\vzero) + g\: a^\dagger(\vzero)\,,
\ee
where $a(\vx)$ is the boson annihilation operator in the position representation at location $\vx$. In yet another notation, 
\be\label{Horig3a}
H_\free = \int d^3\vk \, \Bigl(\tfrac{\hbar^2\vk^2}{2m_y}+E_0\Bigr)\, a^\dagger(\vk) \, a(\vk)
\ee
and
\be\label{Horig3b}
H_\inter= g\int \frac{d^3\vk}{(2\pi)^{3/2}} \, \phi(\vk)
\ee
with $\phi(\vk)$ the bosonic field operator for momentum $\hbar\vk$, i.e.,
\be\label{Horig3c}
\phi(\vk)=a^\dagger(\vk)+a(\vk)
\ee
with $a(\vk)$ the boson annihilation operator for momentum $\hbar\vk$.

Now, for $g\neq 0$, $H_\orig$ as in \eqref{Horig} or \eqref{Horig1b}--\eqref{Horig3b} is UV divergent---it does not actually define an operator in Fock space $\Fock$. Specifically, the integral over $\vk$ in \eqref{Horig3b} diverges for large $|\vk|$ (i.e., for ``ultraviolet'' $\vk$). Equivalently, one can say that the $\delta^3$ that appears as the wave function of the newly created $y$-particle is problematical; it can be thought of as containing an infinite amount of energy. Due to the $\delta^3$ factor, $H_\orig\psi$ is not even a square-integrable function, and thus does not lie in $\Fock$, even if $\psi$ is as nice a function as one could wish for, say, infinitely differentiable and compactly supported in each $(\RRR^3)^n$. Similar divergence problems have plagued most QFTs since the early days of quantum theory \cite{Opp30}. Over the next pages, I will describe how, by means of boundary conditions on $\psi$, it is nevertheless possible to make sense of the formula \eqref{Horig} or \eqref{Horig1b}--\eqref{Horig3b}, that is, how a Hamiltonian can indeed be defined that describes the emission and absorption of particles at a single point $\vzero$. 

Here, by the ``boundary'' of the $n$-particle sector $\Q^{(n)}=(\RRR^3)^n$ we mean the set
\be\label{bou1}
\partial \Q^{(n)} = \Bigl\{ (\vy_1...\vy_{n})\in \Q^{(n)}: \vy_k=\vzero \text{ for some }k \Bigr\}
\ee
of configurations for which at least one $y$-particle meets the $x$-particle. Correspondingly, by the boundary of $\Q$ we mean
\be\label{bou2}
\partial\Q = \bigcup_{n=0}^\infty \partial \Q^{(n)}\,.
\ee
The relevant boundary condition is essentially a relation between the values of $\psi$ at two configurations related through the emission or absorption of a $y$-particle, i.e., at $(\vy_1...\vy_n)\in \Q^{(n)}$ and $(\vy_1...\vy_n,\vzero)\in\partial\Q^{(n+1)}$ or a permutation thereof. Since the latter point lies on the boundary and the former does not, it is called an interior-boundary condition (IBC). The condition is essentially of the form
\be\label{ibc0}
\psi^{(n+1)}(\vy_1...\vy_n,\vzero)= \alpha_n \:\psi^{(n)}(\vy_1...\vy_n)
\ee
with some fixed constant $\alpha_n\in\RRR$. The exact condition, Equation~\eqref{ibc} below, will be a little more involved as the left-hand side needs to be replaced by the leading coefficient in the asymptotics when approaching $(\vy_1...\vy_n,\vzero)$.

\subsection{Motivation: Probability Transport}
\label{sec:transport}

Before I write down the condition in Section~\ref{sec:ibc}, I give a motivation based on a consideration of probability transport. As in $n$-particle quantum mechanics, $|\psi(\vy_1...\vy_n)|^2$ yields the probability density $\rho(\vy_1...\vy_n)$, relative to the $3n$-dimensional volume $d^3\vy_1\cdots d^3\vy_n$, of the configuration being at $(\vy_1...\vy_n)$. Since this is a probability density on $\Q$, its integral over $\Q$ will equal unity,
\be
\int\limits_\Q dy \:|\psi(y)|^2 = \sum_{n=0}^\infty \int\limits_{~\RRR^{3n}} \! d^3\vy_1\cdots d^3\vy_n \: |\psi(\vy_1...\vy_n)|^2 = 1\,.
\ee
And as in quantum mechanics, the transport of probability within each sector $\Q^{(n)}=\RRR^{3n}$ is governed the \emph{probability current} $j$, which is a vector field on $\Q$ given on any sector $\RRR^{3n}$ by
\be\label{jdef}
j = \tfrac{\hbar}{m_y} \Im [\psi^* \nabla \psi]\,.
\ee
However, while in quantum mechanics of $N$ particles, $\rho$ is related to $j$ by the continuity equation 
\be\label{continuity}
\frac{\partial\rho}{\partial t} = -\nabla \cdot j
\ee
in $\RRR^{3N}$, the equation in $\Q$ that looks like \eqref{continuity} cannot be expected to be valid: after all, $j$ only represents probability transport \emph{within one sector} and thus cannot account for probability transport from one sector to another as occurs through particle creation or annihilation. Since particle creation transports probability from whichever configuration $(\vy_1...\vy_n)$ to the configuration $(\vy_1...\vy_n,\vzero)$ (or a permutation thereof), amounts of probability from $\Q^{(n)}$, when transported to $\Q^{(n+1)}$, will there show up at $\partial\Q^{(n+1)}$. And conversely, amounts of probability can be transported from $\Q^{(n+1)}$ to $\Q^{(n)}$ through annihilation, in which case they get transported from $(\vy_1...\vy_n,\vzero)$ (or a permutation thereof) to $(\vy_1...\vy_n)$, so that they can only be transported from $\partial \Q^{(n+1)}$ to $\Q^{(n)}$. So what is the equation in $\Q$ replacing \eqref{continuity}? At configurations $(\vy_1...\vy_n)$ not on the boundary $\partial\Q$, it reads
\be\label{continuity2}
\frac{\partial\rho}{\partial t} = - \nabla \cdot j - J 
\ee
with $J(\vy_1...\vy_n)$ the probability current coming out of $\partial\Q^{(n+1)}$ at $(\vy_1...\vy_n,\vzero)$ (and permutations thereof) minus the probability current flowing into $\partial\Q^{(n+1)}$ at $(\vy_1...\vy_n,\vzero)$ (and permutations thereof). These currents can be expressed as flux integrals of $j$. To see how, let us begin with a simple example: for a current vector field $\vj$ in $\RRR^3$, the flux out of the sphere of radius $r$ minus the flux into it is given by
\be
\int\limits_{\SSS^2}d^2\vomega \: r^2 \vomega \cdot \vj(r\vomega)\,,
\ee
where $\SSS^2$ denotes the unit sphere in $\RRR^3$, and $d^2\vomega$ is the area of a surface element of the sphere. 
Note that $\vomega\cdot \vj$ is the radial component of $\vj$, which is positive for an outward current and negative for inward.
Taking the limit $r\searrow 0$, we obtain the flux out of the origin minus the flux into it. Correspondingly, $J$, the flux out of (permutations of) $(\vy_1...\vy_n,\vzero)$ minus the flux into it, is given by
\be\label{Jdef}
J(\vy_1...\vy_n) = \sum_{k=1}^{n+1} \lim_{r\searrow 0} \int\limits_{\SSS^2} d^2\vomega \: r^2\vomega \cdot \vj_k(\vy_1...\vy_{k-1},r\vomega,\vy_k...\vy_n)\,,
\ee 
where $\vj_k$ means the 3 particle-$k$ components of $j$, i.e., $j=(\vj_1...\vj_{n+1})$. Note that $r\vomega$ does \emph{not} replace $\vy_k$ but gets inserted between $\vy_{k-1}$ and $\vy_k$, so that $\vj_k$ has $n+1$ arguments, and for $k=n+1$ equals $\vj_{n+1}(\vy_1...\vy_n,r\vomega)$.

For the expression \eqref{Jdef} to be non-zero, we need that $\vomega\cdot \vj_k$ should be large, or else the factor $r^2$ in the integral will push the integral to 0 in the limit; that is, $\vomega\cdot \vj_k$ should be large like $1/r^2$. Since $j$ is quadratic in $\psi$, $\psi$ should be large like $1/r$ near $\partial\Q$. 

Now that we know what the needed continuity equation \eqref{continuity2} looks like, we ask what the Hamiltonian should look like in order to entail this continuity equation. Writing
\be
H=H_\free + H_\inter
\ee
with $H_\free=d\Gamma(-\tfrac{\hbar^2}{2m_y}\nabla^2 +E_0)$ (i.e., acting like the first line of \eqref{Horig}), it follows from the Schr\"odinger equation
\be
i\hbar\frac{\partial\psi}{\partial t} = H\psi
\ee
that
\begin{align}
\frac{\partial|\psi|^2}{\partial t}
&= \tfrac{2}{\hbar}\Im[\psi^* (H\psi)]\\
&= \tfrac{2}{\hbar}\Im[\psi^* (H_\free\psi)]
+ \tfrac{2}{\hbar}\Im[\psi^* (H_\inter\psi)]\,.
\end{align}
Since
\be
\tfrac{2}{\hbar}\Im[\psi^* (H_\free\psi)] = -\nabla\cdot j\,,
\ee
we obtain that
\be
\frac{\partial \rho}{\partial t}
= -\nabla\cdot j + \tfrac{2}{\hbar}\Im[\psi^* (H_\inter\psi)]\,,
\ee
and in order to reach the desired form \eqref{continuity2} we need that
\begin{align}
&\tfrac{2}{\hbar}\Im\Bigl[\psi^{(n)}(\vy_1...\vy_n)^*\, \bigl(H_\inter\psi\bigr)^{(n)}(\vy_1...\vy_n) \Bigr] \nonumber\\
&\quad = -J(\vy_1...\vy_n)\label{17}\\[2mm]
&\quad \stackrel{\eqref{Jdef}}{=} -(n+1) \lim_{r\searrow 0} \int\limits_{\SSS^2} d^2\vomega \: r^2\vomega \cdot \vj_{n+1}(\vy_1...\vy_n,r\vomega)\label{18}\\
&\quad \stackrel{\eqref{jdef}}{=} -(n+1) \lim_{r\searrow 0} \int\limits_{\SSS^2} d^2\vomega \: r^2 \tfrac{\hbar}{m_y} \Im\Bigl[\psi^{(n+1)}(\vy_1...\vy_n,r\vomega)^* \:\partial_r \psi^{(n+1)}(\vy_1...\vy_n,r\vomega)\Bigr]\label{19}\\
&\quad = -\tfrac{\hbar}{m_y}(n+1)\: \Im\Biggl[\lim_{r\searrow 0} \int\limits_{\SSS^2} d^2\vomega \: r  \psi^{(n+1)}(\vy_1...\vy_n,r\vomega)^* \:\partial_r (r\psi)^{(n+1)}(\vy_1...\vy_n,r\vomega)\Biggr].\label{20}
\end{align}
In step \eqref{18} we have used that, by permutation symmetry of $\psi$, all $n+1$ summands of \eqref{Jdef} are equal, in step \eqref{19} that $\vomega\cdot\nabla \psi(r\vomega)=\partial_r \psi(r\vomega)$, and in step \eqref{20} that $\partial_r(r\psi)=\psi+r\partial_r\psi$, so that
\be
\Im\bigl[r\psi^* \partial_r(r\psi)\bigr]=\Im\bigl[r\psi^*\psi + r^2\psi^*\partial_r\psi\bigr]= \Im\bigl[r^2\psi^*\partial_r\psi\bigr].
\ee

Now we compare the left-hand side of \eqref{17} to the right-hand side of \eqref{20}. We have imaginary parts on both sides, so the equation would be fulfilled if the square brackets could be made to agree up to the appropriate pre-factors. $(H_\inter\psi)^{(n)}$ could certainly involve $\psi^{(n+1)}$, but the factors of $\psi^*$ seem more problematical. We would need that $r\psi^{(n+1)}(\vy_1...\vy_n,r\vomega)$ is, in the limit $r\to 0$, proportional to $\psi^{(n)}(\vy_1...\vy_n)$. We would need, in other words, a boundary condition on $\psi^{(n+1)}$. We can now easily guess what the IBC should say, except for the choice of the proportionality factor $\alpha_n$, which will become clear soon.

\subsection{Interior-Boundary Condition}
\label{sec:ibc}

Here is what the IBC says: For every $\vomega\in\SSS^2$, $n\in\{0,1,2,\ldots\}$, and $\vy_1...\vy_n\in\RRR^3\setminus\{\vzero\}$,
\be\label{ibc}
\lim_{r\searrow 0} \biggl(r\psi^{(n+1)}(\vy_1...\vy_n,r\vomega)  \biggr) 
= - \frac{g\, m_y}{2\pi\hbar^2\sqrt{n+1}}\: \psi^{(n)}(\vy_1...\vy_n)\,.
\ee
The Hamiltonian $H=H_\ibc$ acts on wave functions satisfying this condition according to
\begin{align}
(H_\ibc\psi)^{(n)}(\vy_1...\vy_n) 
&= -\frac{\hbar^2}{2m_y} \sum_{k=1}^{n} \nabla^2_{k}\psi^{(n)}(\vy_1...\vy_n)+ nE_0\: \psi^{(n)}(\vy_1...\vy_n) \nonumber\\[2mm]
&\quad +\: \frac{g\sqrt{n+1}}{4\pi}\int\limits_{\SSS^2} d^2\vomega \, \lim_{r\searrow 0} \frac{\partial}{\partial r} \Bigl( r \psi^{(n+1)}(\vy_1...\vy_n,r\vomega) \Bigr)\nonumber\\
&\quad +\: \frac{g}{\sqrt{n}} \sum_{k=1}^n  \delta^3(\vy_k)\,\psi^{(n-1)}(\vy_1...\vy_{k-1},\vy_{k+1}...\vy_n)\,.\label{Hibc}
\end{align}
It is now easy to check, by inserting the IBC \eqref{ibc} into \eqref{20}, that
\begin{align}
\eqref{20}
&= \frac{g}{2\pi\hbar}\sqrt{n+1}\: \Im\Biggl[\lim_{r\searrow 0} \int\limits_{\SSS^2} d^2\vomega \: \psi^{(n)}(\vy_1...\vy_n)^* \:\partial_r (r\psi)^{(n+1)}(\vy_1...\vy_n,r\vomega)\Biggr]\\
&= \frac{2}{\hbar}\: \Im\Biggl[\psi^{(n)}(\vy_1...\vy_n)^* \: \frac{g\sqrt{n+1}}{4\pi}\int\limits_{\SSS^2} d^2\vomega \: \lim_{r\searrow 0} \partial_r (r\psi)^{(n+1)}(\vy_1...\vy_n,r\vomega)\Biggr]\,,
\end{align}
in agreement with the left-hand side of \eqref{17} and \eqref{Hibc} at a configuration not on the boundary.

Equations \eqref{ibc} and \eqref{Hibc} form the solution to the UV divergence problem that I want to present: 

\bigskip

\noindent{\bf Theorem.}~\cite{LSTT17} \emph{Equation~\eqref{Hibc} provides a well-defined, self-adjoint operator $H_\ibc$ on a dense domain in Fock space $\Fock$ containing wave functions $\psi$ that satisfy the IBC \eqref{ibc}. The operator is bounded from below.} 

\bigskip

This means that the UV divergence is absent from $H_\ibc$.
And I argue below that $H_\ibc$ is a reasonable interpretation of the original expression $H_\orig$.

Sometimes in QFT, one uses boundary conditions on 1-particle propagators, i.e., on Green's functions. The boundary condition \eqref{ibc} is not like that. For one thing, it is a condition on the wave function, but more importantly, it is not a 1-particle condition but instead links the $n$-particle sector with the $n+1$-particle sector.

\subsection{Why It Works, and Why It Is Reasonable}

Two things may seem surprising: First, how the divergence problem can be absent if the $\delta^3$ that was the root of the trouble is still present. And second, why anyone would regard the middle line of \eqref{Hibc} as a reasonable interpretation of the middle line of \eqref{Horig}.

Here is why $\delta^3$ does not cause trouble any more: Note that $\psi^{(n)}$ diverges at the boundary $\partial\Q^{(n)}$ like $1/r$ with $r$ the distance from the boundary,
\be\label{asymp1}
\psi^{(n)}(\vy_1...\vy_{n-1},r\vomega) \approx c \: r^{-1}
\ee
with a complex coefficient $c$ that will depend on $\vy_1,\ldots,\vy_{n-1}$. Note further that the Laplacian of $1/r$ is a delta function,
\be\label{LaplaceCoulomb}
\nabla^2_{\vy} \: \frac{1}{|\vy|} = -4\pi\delta^3(\vy),
\ee
a fact familiar from electrostatics, where the potential $\phi$ obeys the Poisson equation $\nabla^2\phi=-4\pi\rho$ with the charge density $\rho$, and the potential of a point charge $\rho=\delta^3$ is the Coulomb potential $1/r$. As a consequence, the Laplacian of $\psi^{(n)}$ will contain $\delta^3$ contributions, and they cancel exactly the $\delta^3$ contributions in the definition \eqref{Hibc} of $H_\ibc$, with the effect that $H_\ibc\psi$ is actually a square-integrable function and thus lies in $\Fock$. Indeed, putting \eqref{asymp1} and \eqref{LaplaceCoulomb} together, we obtain that
\be
-\tfrac{\hbar^2}{2m_y}\nabla_{n}^2 \psi^{(n)}(\vy_1...\vy_n) = c \tfrac{4\pi\hbar^2}{2m_y} \delta^3(\vy_n) + \text{a function}\,,
\ee
where we have separated the singular (distributional) part (that involves a Dirac delta) from the regular part (that is a function). By virtue of the IBC \eqref{ibc},
\be
c= - \frac{g\, m_y}{2\pi\hbar^2\sqrt{n}}\: \psi^{(n-1)}(\vy_1...\vy_{n-1})\,,
\ee
so
\be
-\tfrac{\hbar^2}{2m_y}\nabla_{n}^2 \psi^{(n)}(\vy_1...\vy_n) = - \frac{g}{\sqrt{n}} \delta^3(\vy_n)\: \psi^{(n-1)}(\vy_1...\vy_{n-1}) + \text{a function}\,,
\ee
which cancels exactly the $\delta^3$ in the $n$-th summand of the last line of \eqref{Hibc}.

Let us turn to a comparison between the middle line of \eqref{Hibc} and that of \eqref{Horig}. Suppose for a moment (I will relax this supposition by the end of the subsection) that $\psi^{(n+1)}(\vy_1...\vy_n,r\vomega)$ can be expanded into powers of $r$,
\be\label{powersofr}
\psi^{(n+1)}(\vy_1...\vy_n,r\vomega) = \sum_{\ell=-1}^\infty c_\ell \, r^\ell
\ee
starting at exponent $-1$, with complex coefficients $c_\ell$ that may depend on $\vy_1,\ldots,\vy_n$ and $\vomega$. Then
\begin{align}
\partial_r\Bigl(r\psi^{(n+1)}(\vy_1...\vy_n,r\vomega)\Bigr)
& =\partial_r\Bigl(r\sum_{\ell=-1}^\infty c_\ell \, r^\ell\Bigr)\\
& =\partial_r\sum_{\ell=-1}^\infty c_\ell \, r^{\ell+1}\\
& =\sum_{\ell=0}^\infty c_\ell \, (\ell+1) \, r^{\ell}
\end{align}
starting at exponent 0, and taking $r\to0$, only the $\ell=0$ term remains, yielding $c_0$. Thus, the middle line of $H_\ibc$ in \eqref{Hibc} equals
\be\label{c0}
\frac{g\sqrt{n+1}}{4\pi} \int\limits_{\SSS^2} d^2\vomega\: c_0(\vy_1...\vy_n,\vomega)\,.
\ee
For comparison, the middle line of $H_\orig$ in \eqref{Horig} asks us to evaluate $\psi^{(n+1)}$ at $(\vy_1...\vy_n,\vzero)$; since $\psi^{(n+1)}$ diverges there, reasonable interpretations could be: either to take the leading coefficient $c_{-1}$, or to use the non-divergent part $\sum_{\ell=0}^\infty c_\ell \, r^\ell$ and evaluate \emph{that} at $r=0$, which yields $c_0$. Since the IBC \eqref{ibc} fixes $c_{-1}$, as it demands that
\be
c_{-1}(\vy_1...\vy_n,\vomega) = -\frac{g\, m_y}{2\pi\hbar^2\sqrt{n+1}}\: \psi^{(n)}(\vy_1...\vy_n)\,,
\ee
this first interpretation would not actually couple $H\psi^{(n)}$ to $\psi^{(n+1)}$ and therefore would not work. That leaves us with the second interpretation, according to which the middle line of \eqref{Horig} means $g\sqrt{n+1}\, c_0(\vy_1...\vy_n,\vomega)$. Here, the question arises which $\vomega$ to use, and an obvious choice is to average over all $\vomega$s, which leads us to \eqref{c0}, which coincides with the middle line of $H_\ibc$ in \eqref{Hibc}. By the way, it turns out that for $\psi$ in the domain of $H_\ibc$, both $c_{-1}$ and $c_0$ are actually independent of $\vomega$ (or else $\nabla^2\psi$ would not be square-integrable away from the boundary, see \cite[Sec.~3.3 Rem.~7]{bohmibc}), so that averaging over $\vomega$ becomes unnecessary.

As a technical remark, we can drop the supposition that $\psi^{(n+1)}(\vy_1...\vy_n,r\vomega)$ can be expanded into powers of $r$ as in \eqref{powersofr}. As proved in \cite{LSTT17}, every $\psi$ in the domain of $H_\ibc$ is, as $r\to 0$, of the asymptotic form
\be
\psi^{(n+1)}(\vy_1...\vy_n,r\vomega) 
= c_{-1}(\vy_1...\vy_n)\,r^{-1} + f(\vy_1...\vy_n,r\vomega)
\ee
with a function $f$ from the second Sobolev space; for such $f$, both $f$ and $\partial_r f$ possess a finite limit as $r\to 0$, so that
\be
\lim_{r\to0}\partial_r\Bigl(r\psi^{(n+1)}(\vy_1...\vy_n,r\vomega) \Bigr)
= f(\vy_1...\vy_n,\vzero)\,,
\ee
which is the analog of the $c_0(\vy_1...\vy_n)$ mentioned before and leads us again to the conclusion that the middle line of $H_\ibc$ in \eqref{Hibc} agrees with a reasonable interpretation of the middle line of $H_\orig$ in \eqref{Horig}.

\subsection{Comparison to Renormalization}
\label{sec:renormalization}

Another argument to the effect that $H_\ibc$ is a reasonable interpretation of $H_\orig$ comes from renormalization. Here, renormalization means to introduce a UV cut-off into the Hamiltonian and then take a limit of removing the cut-off, possibly after adding a counter-term, in order to obtain a limiting Hamiltonian $H_\infty$ called the renormalized Hamiltonian. For the model discussed above, it has long been known that such a limit $H_\infty$ exists \cite{vH52,Der03}, and it has recently turned out \cite{LSTT17} that $H_\infty$ agrees with $H_\ibc$ up to addition of a constant (i.e., of a multiple of the identity),
\be
H_\infty= H_\ibc +E
\ee
for some $E\in\RRR$. Needless to say, Hamiltonians that differ only by addition of a constant $E$, $H'=H+E$, are usually regarded as physically equivalent because the time-evolved state vectors they generate differ only by a global phase factor,
\be
e^{-iH't/\hbar}\psi_0= e^{-iEt/\hbar} e^{-iHt/\hbar}\psi_0\,,
\ee
and are thus considered physically equivalent.
The fact that two different approaches to obtaining a well-defined Hamiltonian from the expression $H_\orig$ lead to the same result suggests that this result is indeed a reasonable, physically appropriate interpretation of $H_\orig$.

In more detail, to introduce a cut-off means here to replace the delta distribution $\delta^3$ in $H_\orig$ by a regular function $\varphi:\RRR^3\to\RRR$ approximating it. We may assume that $\varphi$ is smooth, compactly supported, thus square-integrable, and perhaps rotationally symmetric. The cut-off Hamiltonian is then given by
\begin{align}\label{Hphi}
(H_\varphi\psi)^{(n)}(\vy_1...\vy_n)
&= -\frac{\hbar^2}{2m_y} \sum_{k=1}^n \nabla_k^2 \psi^{(n)}(\vy_1...\vy_n) + n E_0\: \psi^{(n)}(\vy_1...\vy_n)\nonumber\\
&\quad + g\, \sqrt{n+1}\int_{\RRR^3} \!\! d^3\vy\: \varphi(\vy) \: \psi^{(n+1)}(\vy_1...\vy_n,\vy)\nonumber\\
&\quad + \frac{g}{\sqrt{n}} \sum_{k=1}^n \varphi(\vy_k) \: \psi^{(n-1)}(\vy_1...\vy_{k-1},\vy_{k+1}...\vy_n)\,.
\end{align}
Here, $\delta^3$ was replaced by $\varphi$ in the last line, and evaluation at $\vzero$ was replaced by the inner product with $\varphi$ in the last variable in the middle line. $H_\varphi$ is known (e.g., \cite{Der03}) to be well-defined and self-adjoint. One may think of $\varphi$ as a continuous charge distribution, with the effect that the $x$-particle is no longer a point particle but spread out and can emit and absorb $y$-particles not only at the origin $\vzero$ but in a whole neighborhood.

To remove the cut-off then means to take a limit in which $\varphi$ converges to $\delta^3$ (denoted $\varphi \to \delta^3$). As I will outline at the end of this subsection, it can be shown \cite{vH52,Der03,LSTT17} that, if the rest energy $E_0$ in \eqref{Hphi} is positive, then there exists a self-adjoint operator $H_\infty$ and for every $\varphi$ a real number $E_\varphi$ such that $E_\varphi\to \infty$ and
\be
H_\varphi + E_\varphi \to H_\infty ~~~\text{as}~~\varphi\to\delta^3.
\ee
Remarkably, although the existence of $H_\infty$ has long been known, it had not been realized until recently that the functions in the domain of $H_\infty$ satisfy the IBC \eqref{ibc}, and that $H_\infty$ acts on them according to the formula \eqref{Hibc} up to addition of a constant. 

The trick for taking the limit $\varphi \to \delta^3$ is \cite{Der03} to write $H_\varphi$, which is 
\be
H_\varphi = \int d^3\vk\biggl( \omega(\vk)\, a^\dagger(\vk) \, a(\vk) + g\, \hat\varphi(\vk)^* a(\vk) + g\, \hat\varphi(\vk) \, a^\dagger(\vk)\biggr)
\ee
with $\omega(\vk)=\hbar^2\vk^2/2m_y + E_0$, in the form of completing the square,
\be
H_\varphi = \int d^3\vk\, \omega(\vk)\biggl(a^\dagger(\vk) + g\frac{\hat\varphi(\vk)^*}{\omega(\vk)}\biggr) \biggl( a(\vk) + g \frac{\hat\varphi(\vk)}{\omega(\vk)}\biggr)-\int d^3\vk\,  g^2\frac{\bigl|\hat\varphi(\vk)\bigr|^2}{\omega(\vk)}\,,
\ee
and regard the last integral as $E_\varphi$. The unitary Bogolyubov transformation (dressing operator)
\be\label{Wdef}
W_\varphi := \exp\biggl( \int d^3\vk \, \frac{\hat\varphi(\vk)^*}{\omega(\vk)}a(\vk) - \int d^3\vk \, \frac{\hat\varphi(\vk)}{\omega(\vk)}a^\dagger(\vk)\biggr)
\ee
intertwines $H_\varphi+E_\varphi$ with the free Hamiltonian,
\be
H_\varphi+E_\varphi = W_\varphi\, H_\free\, W_\varphi^\dagger,
\ee
and for $E_0>0$ possesses a limit $W_\varphi\to W_\infty$ as $\hat\varphi(\vk)\to \mathrm{const.}$ or $\varphi\to\delta^3$, as already suggested by the fact that $1/\omega(\vk)$ is still square integrable, so that the right-hand side of \eqref{Wdef} is well defined when replacing $\hat\varphi$ by a real constant. Putting the pieces together, $H_\varphi+E_\varphi \to H_\infty =W_\infty\, H_\free\, W_\infty^\dagger$.

As a future project, it would also be of interest to carry out a comparison of the IBC approach with the Epstein--Glaser approach \cite{EG73,Sch95} to renormalization of the perturbation series for the scattering matrix.

\subsection{Other Models, and Literature}
\label{sec:literature}

The mathematical analysis of IBC Hamiltonians is similar to that of 
point interaction (zero-range interaction) \cite{DFT08}, i.e., potentials involving Dirac delta distributions as in
\be
H\psi(\vx)=-\nabla^2\psi + g\, \delta^3(\vx) \: \psi(\vx)\,.
\ee
In fact, also point interaction can be expressed through a boundary condition on the wave function, the Bethe-Peierls boundary condition \cite{BP35}
\be\label{BethePeierls}
\lim_{r\searrow 0}\Bigl(\alpha +\frac{\partial}{\partial r} \Bigr) \Bigl(r\psi(r\vomega)\Bigr) =0
\ee
with given constant $\alpha\in\RRR$.

The IBC \eqref{ibc} is not the only possible IBC for an $x$-particle fixed at the origin. Rather, there is a 4-parameter family of IBCs and associated self-adjoint Hamiltonians \cite{Yaf92,TT15a,TT15b,LSTT17}, which can be thought of as having, in addition to the particle creation and annihilation at the origin, also a point interaction at the origin \cite{LSTT17,ST18}. These Hamiltonians and IBCs can be characterized as follows. On wave functions that diverge on the boundary like $1/r$, we can define two annihilation operators, $A$ and $B$, that extract from $\psi$ as in \eqref{powersofr} the coefficient $c_0$ of order 0 and $c_{-1}$ of order $-1$, respectively:
\be
A\psi^{(n)}(\vy_1...\vy_n)= \sqrt{n+1} \: c_0(\vy_1...\vy_n)\,,~~~
B\psi^{(n)}(\vy_1...\vy_n)= \sqrt{n+1} \: c_{-1}(\vy_1...\vy_n)\,.
\ee
Then the IBC \eqref{ibc} can be rewritten as
\be
B\psi = -\frac{g\, m_y}{2\pi\hbar^2}\: \psi
\ee
and the action of the Hamiltonian $H$ \eqref{Hibc} as
\be
H_\ibc = H_\free + g A + ga^\dagger(\vzero)\,.
\ee
Now replacing $A\to e^{i\theta}(\alpha A+\beta B)$ and $B\to e^{i\theta}(\gamma A +\delta B)$ with real constants $\alpha,\beta,\gamma,\delta,\theta$ subject to $\alpha\delta-\beta\gamma =1$ yields the further members of the 4-parameter family of IBCs and self-adjoint Hamiltonians \cite[Sec.~4]{LSTT17}. The families of possible IBCs for higher spins were identified in \cite{co1}.

While the model of Sections \ref{sec:model}--\ref{sec:renormalization} above assumes that the $x$-particle is fixed at the origin, a more realistic model assumes several moving $x$-particles. The Hilbert space is then of the form $\Hilbert=\Fock_x \otimes \Fock_y$ with $\Fock_x$ a fermionic and $\Fock_y$ a bosonic Fock space, and the configuration space is of the form $\Q_x \times \Q_y$ with each factor a copy of the $\Q$ as in \eqref{Qdef}. The relevant boundary is then the set
\be\label{bou3}
\partial\Q = \Bigl\{ (\vx_1...\vx_m,\vy_1...\vy_n)\in \Q_x\times \Q_y: \vx_j=\vy_k\text{ for some }j,k \Bigr\}
\ee 
of ``collision configurations'' (i.e., ones in which an $x$ and a $y$ meet). As shown by Lampart~\cite{Lam18}, there exists a well-defined, self-adjoint IBC Hamiltonian for this model. This is particularly remarkable insofar as no renormalized Hamiltonian $H_\infty$ was known for this case prior to Lampart's work. Moreover, since in this model the formal expression for the interaction Hamiltonian is, instead of \eqref{Horig2b},
\be
H_\inter = g\int d^3\vx \: b^\dagger(\vx)\, \bigl( a(\vx) + a^\dagger(\vx)\bigr)\, b(\vx)
\ee
with $b$ the fermion annihilation operator, the model is also an example in which the Hamiltonian is not quadratic in the creation and annihilation operators.

IBCs have been studied mathematically for several other cases: Lampart and Schmidt \cite{LS18} proved the existence of an IBC Hamiltonian for a model with moving $x$-particles in 2 dimensions, as well as for a model with a different dispersion relation, replacing the Laplace operator $-\tfrac{\hbar^2}{2m_y}\nabla^2$ by the pseudo-relativistic 1-particle Hamiltonian $\sqrt{m_y^2 - \nabla^2}$, a model for which Nelson~\cite{Nel64} could prove the existence of a renormalized Hamiltonian in 1964. IBC models in 1 dimension were studied in \cite{KS15,LN18}. Early works on the model with $x$ fixed at $\vzero$~\cite{Tho84,Yaf92} used a truncated Fock space with only 2 or 3 sectors.

Other profiles than $\delta^3$ for the emission and absorption terms are often desired in connection with relativistic dispersion relations; they lead to conditions that are no longer literally boundary conditions but have been termed ``abstract boundary conditions''~\cite{BM14}; the IBC approach has been extended in this direction in \cite{LS18,Sch18,Sch19,Lam20,Pos20}.

The considerations of probability transport in Section~\ref{sec:transport} apply literally in the Bohmian picture, in which the particles are attributed trajectories that can begin and end at emission and absorption events, corresponding to jumps of the actual configuration from $\Q^{(n)}$ to $\partial\Q^{(n+1)}$ or vice versa \cite{Tum04,bohmibc}. 

A pedagogical introduction to IBCs can be found in \cite{TT15b}; the behavior of IBCs under time reversal was studied in \cite{ST18}. Apart from the boundary sets \eqref{bou1} and \eqref{bou3}, it is natural to also consider other domains with boundary, and among them in particular with boundaries of codimension 1 (whereas \eqref{bou1} and \eqref{bou3} have codimension 3). First discussed in \cite{Tum04,TT15b}, IBCs on codimension-1 boundaries were systematically studied in \cite{co1} for the Laplacian (i.e., for non-relativistic Hamiltonians) and in \cite{LN18,IBCdiracCo1} for the Dirac equation. Lienert and Nickel~\cite{LN18} developed a QFT model in 1 space dimension in which moving $x$-particles emit and absorb other $x$s, $x\rightleftarrows 2x$, based on the Dirac operator as the free Hamiltonian; they specified an IBC and proved results implying the existence of a well-defined, self-adjoint Hamiltonian. 

In 3 space dimensions, in contrast, the free Dirac operator does not allow for a boundary condition on the set of collision configurations (or any subset of codimension 3) and thus not for an IBC \cite{HT20}. That sounds discouraging at first. However, in nature the relevant particle reaction is not $x\rightleftarrows 2x$ but $x\rightleftarrows x+y$ with the $y$-particles being photons, governed not by the Dirac equation but by the Maxwell equation, and this case has not been settled yet. Moreover, despite the impossibility theorem just described, a model in which Dirac particles get emitted and absorbed by a spin-0 particle fixed at the origin has been proved \cite{HT20} to possess a well-defined and self-adjoint IBC Hamiltonian if the Dirac particles feel a sufficiently strong $1/r$ potential (such as a Coulomb or gravitational potential), regardless of whether the potential is attractive or repulsive. It would be of interest to study rigorously whether an IBC Hamiltonian exists for Dirac particles in a general-relativistic gravitational field of a point mass (such as the Reissner-Nordstr\"om space-time geometry); preliminary (non-rigorous) considerations in this direction can be found in \cite{timelike}.

\subsection{Interior-Boundary Conditions for the Dirac Equation}

Let me give some explicit examples of what IBCs look like for the Dirac equation. I will stay away from the difficulties and technicalities associated with the situation in which the boundary has codimension 3, a situation that arises in 3 space dimensions when the boundary is the set of configurations in which two particles meet, as in \eqref{bou1} and \eqref{bou3}. Instead, I will consider the more elementary example of a boundary of codimension 1 \cite{LN18,IBCdiracCo1}, which arises either when we consider space dimension 1 or when we study an artificially simplified example of a boundary. Here, we will do the latter.

So consider, as a simple example, the free 1-particle Dirac equation 
\be\label{Dirac1}
i\hbar \partial_t \psi= -i\hbar\valpha\cdot\nabla \psi +m\beta \psi
\ee
in 3 space dimensions in a region $\Omega\subset \RRR^3$ with 2d boundary surface $\partial \Omega$. For concreteness, let us consider the upper half space $\Omega=\{(x_1,x_2,x_3):x_3\geq 0\}$ with boundary $\partial \Omega$ the $x_1x_2$-plane. 

Boundary conditions for the Dirac equation specify two out of the four components of $\psi$ at every boundary point and leave the other two components arbitrary. More precisely, they split the Dirac spin space $S=\CCC^4$ into a direct sum of two 2d subspaces, $S=U\oplus V$, specify the part $\psi_U$ of $\psi=\psi_U+\psi_V$ in $U$ and put no condition on the other part $\psi_V$. 

For example, let us look at the best known case of a \emph{reflecting} boundary condition, viz., the boundary condition of the ``MIT bag model'' of quark confinement \cite{CJJTW74},
\be\label{MIT}
(\gamma^3 -i)\psi(x_1,x_2,0) =0\,.
\ee
(Here, $i$ means $i$ times the identity matrix.)
Since $\gamma^3$ is unitarily diagonalizable with eigenvalues $\pm i$ and 2d eigenspaces, $\gamma^3-i$ is $-2i$ times an orthogonal projection; so this condition says that, at every boundary point, $\psi$ has to lie in the eigenspace with eigenvalue $+i$; put differently, $U$ is the eigenspace with eigenvalue $-i$, $\psi_U$ is prescribed to vanish, $V$ is the eigenspace with eigenvalue $+i$, and no condition is put on $\psi_V$. 

If the situation that half of the components can be prescribed seems unfamiliar, then it may be useful to note that in the non-relativistic case, a boundary condition may involve, at every boundary point $\vx\in\partial \Omega$, the value $\psi(\vx)$ as well as the normal derivative $\partial_n\psi(\vx)$ to the boundary at $\vx$. If $\psi$ has $s$ components, then $(\psi(\vx),\partial_n\psi(\vx))$ has $2s$ components, and the boundary condition will usually be of the form
\be\label{Robin}
A(\vx)\, \psi(\vx) + B(\vx) \, \partial_n\psi(\vx)=0
\ee
with $s\times s$ matrices $A$ and $B$. So the condition,
consisting of $s$ equations, specifies half of the components while putting no condition on the other half (more precisely, it specifies the part of $(\psi(\vx),\partial_n\psi(\vx))$ in an $s$-dimensional subspace of $\CCC^{2s}$, that is, in half of the dimensions). 

Now I will describe an example of an IBC analogous to \eqref{MIT}. For such an example, we need at least two sectors of the wave function and of configuration space. So let us assume, for simplicity, that there are exactly two sectors;
that the configuration space is $\Q=\Q^{(0)} \cup \Q^{(1)}$, where $\Q^{(0)}$ has just one element (the ``empty configuration'') and $\Q^{(1)}=\Omega$; and that the Hilbert space is $\Hilbert=\CCC\oplus L^2(\Omega,\CCC^4)$.
The function $\psi^{(1)}$ obeys the Dirac equation \eqref{Dirac1} at every point in the half space $\Omega$, while 
\be\label{ex1psi0}
i\hbar \frac{\partial \psi^{(0)}}{\partial t} = \int_{\RRR^2} \!\! dx_1\, dx_2\, N(x_1,x_2)^\dagger\, \psi^{(1)}(x_1,x_2,0)\,,
\ee
where $N(x_1,x_2)$ is a fixed spinor field that is square-integrable and satisfies
\be\label{Nalpha}
N^\dagger(x_1,x_2) \, \gamma^3\gamma^0 \, N(x_1,x_2) =0
\ee
at every $(x_1,x_2)\in\RRR^2$. The IBC reads
\be\label{IBCDirac}
(\gamma^3-i) \psi^{(1)}(x_1,x_2,0)
=-\tfrac{i}{\hbar} (\gamma^3-i) \gamma^3 \gamma^0 N(x_1,x_2) \, \psi^{(0)}\,.
\ee
Again, the boundary condition prescribes the part of $\psi^{(1)}(x_1,x_2,0)$ in $U$ while putting no condition on its part in $V$, and again, the condition couples the two sectors of $\psi$. The spinor $N$ remains arbitrary as a parameter of the model. See \cite{IBCdiracCo1} for a proof of self-adjointness of the Hamiltonian defined by \eqref{Dirac1} and \eqref{ex1psi0} on a domain of wave functions satisfying the IBC \eqref{IBCDirac} (in a slightly different setting with compact $\partial \Omega$) and a discussion of further variants of this IBC.

\section{Quantum Electrodynamics}
\label{sec:qed}

IBCs have not yet been made to work for QED, but I want to describe reasons for thinking that they can, and more generally for thinking that wave functions in the particle-position representation can be helpful in QED. To this end, I first need to talk about multi-time wave functions.

\subsection{Multi-Time Wave Functions}

Wave functions in the particle-position representation are functions of the positions of the particles, $\psi=\psi(\vx_1...\vx_n)$. In the non-relativistic picture, this wave function evolves with time. In the relativistic picture, it does not seem appropriate any more to consider several space points at the same time coordinate, as that situation would depend on the choice of Lorentz frame. The natural relativistic version would be to consider wave functions that are functions of several space-time points, $\phi=\phi(x_1...x_n)$ with each $x_j\in \RRR^4$ or $x_j\in\sM$ with $\sM$ the space-time manifold. We usually take $\phi$ to be defined on the set of \emph{spacelike} space-time configurations
\be
\sS=\bigcup_{n=0}^\infty \sS_n = \bigcup_{n=0}^\infty \Bigl\{(x_1...x_n)\in \sM^n: x_j=x_k \text{ or }x_j \spacelike x_k \text{ for all }j,k \Bigr\}\,,
\ee
where $x\spacelike y$ means that $x$ and $y$ are spacelike separated. Since $\phi$ involves several time variables $x_1^0,\ldots,x_n^0$, it is called a multi-time wave function. This concept, a covariant expression of the state vector, was suggested early on in the history of quantum physics by Eddington \cite{Edd1929} and Dirac \cite{dirac:1932}; see \cite{LPT} for comprehensive discussion and further references.

Since $\phi^{(n)}=\phi\big|_{\sS_n}$ depends on $n$ time variables, its time evolution, if governed by PDEs, needs $n$ equations, one for each time variable, as in 
\be\label{multi}
i\hbar \frac{\partial \phi^{(n)}}{\partial x_j^0} = H_j^{(n)} \phi\,,
\ee 
where $H_j^{(n)}$ is an operator called the $j$-th partial Hamiltonian; it can roughly be thought of as collecting the terms in the Hamiltonian pertaining to $x_j$, and we have set $c=1$. Such a system of equations can be inconsistent, but consistency has been verified for relativistic versions of the Lee model analogous to \eqref{Horig} \cite{PT13,LN18,LNT20}. An upshot at this point is that the particle-position representation fits nicely together with Lorentz invariance.

A version of the Born rule appropriate for $\phi$ is the \emph{curved Born rule} \cite{LT:2017,LT:2021b}: \emph{If detectors are placed along a (possibly curved) Cauchy surface $\Sigma$ (i.e., a spacelike 3-surface) in $\sM$, then the probability density (relative to the Riemannian volume measure) of finding the configuration $(x_1,\ldots,x_n)\in\Sigma^n$ is given by}
\be
\rho(x_1...x_n) = |\phi^{(n)}(x_1...x_n)|^2 \,,
\ee
with $|\cdot|^2$ suitably understood (such as,
\be
|\phi(x_1...x_n)|^2=
\overline\phi(x_1...x_n) \Bigl[\gamma^{\mu_1}n_{\mu_1}(x_1)\otimes \cdots \otimes \gamma^{\mu_n}n_{\mu_n}(x_n)\Bigr] \phi(x_1...x_n)
\ee
for Dirac wave functions with $n_\mu(x)$ the future unit normal vector to $\Sigma$ at $x$).

\subsection{Landau and Peierls}
\label{sec:LP}

In 1930, Landau and Peierls \cite{lp:1930} wrote down a version of QED (simplified without positrons) in the particle-position representation. Here I reproduce their equations in a multi-time form developed recently \cite{LT:2021a}. The wave function depends on the space-time points $x_1,\ldots,x_m$ of a variable number $m$ of electrons, as well as the space-time points $y_1,\ldots,y_n$ of a variable number $n$ of photons, $\phi=\phi(x_1...x_m,y_1...y_n)$. We take it that the wave function of a single electron is a Dirac wave function with 4 complex components labeled by the index $s=1...4$, and the wave function of a single photon is a complexified Maxwell field represented by its vector potential $A_\mu$, also with 4 complex components but now labeled by the index $\mu=0...3$. Thus,
\be
\phi=\phi^{(m,n)}_{s_1...s_m\mu_1...\mu_n}(x_1...x_m,y_1...y_n)\,.
\ee
A Lorentz transformation acts on each $x_j$ and each $y_k$ in the usual way, on each index $\mu_k$ as on the index of a world vector, and on each index $s_j$ as on the index of a Dirac spinor; thus, the action on $\phi^{(m,n)}$ is the product of representations of the proper Lorentz group. The domain of definition of $\phi$ is the set of spacelike configurations of $x$s and $y$s, i.e., where any two of $x_1...x_m,y_1...y_n$ are either spacelike or equal. $\phi$ is symmetric against permutation of the photons (permuting the space-time points $y_k$ and the indices $\mu_k$ in the same way) and anti-symmetric against permutation of the electrons (of $x_j$ and $s_j$ alike). Let $e$ and $m_x$ denote the charge and the mass of the electron. The time evolution is governed by 3 equations, \eqref{LPx}, \eqref{LPy}, and \eqref{LPg} below. For now, we will not worry about UV divergence and simply write down Dirac delta distributions.

The equation in $x_j$ is the Dirac equation with an additional term:
\be\label{LPx}
(i\gamma^\mu_j\partial_{x_j,\mu}-m_x)\phi^{(m,n)}(x_1...x_m,y_1...y_n)
= e\sqrt{n+1} \:\gamma^\rho_j\: \phi^{(m,n+1)}_{\mu_{n+1}=\rho}(x_1...x_m,y_1...y_n,x_j)
\ee
Here, $\gamma_j^\mu$ means $\gamma^\mu$ acting on the index $s_j$, and most indices have not been made explicit. 

The equation in $y_k$ is the Maxwell equation with a source term:
\begin{multline}\label{LPy}
2\partial_{y_k}^\mu\partial^{~}_{y_k,[\mu}\phi^{(m,n)}_{\mu_k=\nu]}(x_1...x_m,y_1...y_n) =\\
\frac{4\pi e}{\sqrt{n}}\sum_{j=1}^m \delta^3_{\mu}(y_k-x_j) \: \gamma_j^\mu \gamma_{j\nu} \:\phi^{(m,n-1)}_{\widehat{\mu_k}}(x_1...x_m,y_1...y_{k-1},y_{k+1}...y_n)\,.
\end{multline}
Here, $[\mu\nu]$ means anti-symmetrization in the index pair as in $S_{[\mu\nu]}=\tfrac12(S_{\mu\nu}-S_{\nu\mu})$; the 3-dimensional Dirac delta distribution wears a space-time index $\mu$ because it is a vector field in the 4 dimensional set $\{x\in\RRR^4:x=0 \text{ or }x\spacelike 0\}$ (corresponding to a 3-form that can be integrated over any spacelike surface through $0\in\RRR^4$ \cite{LT:2021a}); and $\widehat{\mu_k}$ means that the index $\mu_k$ is omitted.

The third equation is a gauge condition that is actually not Lorentz invariant,
\be\label{LPg}
\sum_{\mu_k=1}^3 \partial_{y_k}^{\mu_k} \phi^{(m,n)}_{\mu_k}(x_1...x_m,y_1...y_n)=0 \,,
\ee
where $\mu_k=0$ is omitted. Again, most indices are not made explicit. This equation is analogous to the Coulomb gauge condition
\be\label{Coulomb}
\sum_{\mu=1}^3 \partial^{\mu} A_\mu=0\,.
\ee

To understand the equations \eqref{LPx}--\eqref{LPg}, let us compare them to 1-particle equations. The 1-particle Dirac equation in an external electromagnetic field with vector potential $A_\mu(x)$ reads
\be\label{Dirac}
(i\gamma^\mu\partial_{\mu}-m_x)\phi(x)
= e \,\gamma^\rho A_\rho(x) \: \phi(x)\,.
\ee
Equation \eqref{LPx} has the same structure with three differences: it applies to $x_j$ in a multi-time function, the factor $\sqrt{n+1}$ in a sense compensates overcounting due to symmetrization, and, most importantly, the external field $A_\mu$ has been replaced by the wave function of the $n+1$-st photon. If $\phi^{(m,n+1)}$ factorized according to
\be
\phi^{(m,n+1)}_{\mu_{n+1}}(x_1...x_m,y_1...y_{n+1}) = A_{\mu_{n+1}}(y_{n+1}) \: \phi^{(m,n)}(x_1...x_m,y_1...y_n)\,,
\ee 
then \eqref{LPx} would reduce exactly to \eqref{Dirac} applied to $x_j$ (except for the factor $\sqrt{n+1}$ that stems from our normalization convention). In other words, \eqref{LPx} is essentially the Dirac equation in which the vector potential is provided by the wave function of the next photon.

The 1-photon equation, i.e., the complex Maxwell equation, with source term $J_\nu(y)$ reads
\be\label{Maxwell}
2\partial^\mu \partial_{[\mu} A_{\nu]}(y) = 4\pi J_\nu(y)\,,
\ee
which is a re-formulation of the well-known form $\partial^\mu F_{\mu\nu} = 4\pi J_\nu$ using $F_{\mu\nu} = 2 \partial_{[\mu}A_{\nu]}$.
Equation \eqref{LPy} has the same structure, up to the factor $1/\sqrt{n}$ that compensates overcounting, with $\phi_{\mu_k}$ playing the role of $A_\mu$, $y_k$ the role of $y$, and the source term given by
\be
J_\nu(y) = e \sum_{j=1}^m \delta^3_\mu(y-x_j) \:\gamma_j^\mu \, \gamma_{j,\nu} \: \phi^{(m,n-1)}_{\widehat{\mu_k}}(\widehat{y_k})\,.
\ee
This term is indeed concentrated on the locations $x_1,\ldots,x_m$ of the electrons; moreover, when the term is integrated over a horizontal surface with respect to the chosen Lorentz frame, the Dirac delta contributes only for $\mu=0$, so $\gamma^\mu\gamma_\nu$ becomes $\gamma^0\gamma_\nu=\alpha_\nu$, which is the matrix associated with the current in the wave function $\phi$.

What I am getting at is that \eqref{LPx} and \eqref{LPy} are very natural equations; they are basically simply the Dirac and Maxwell equations applied to particular variables of $\phi$, with natural expressions inserted for the vector potential and the source term, given that the photon aspect of $\phi$ is analogous to $A_\mu$ and the electron aspect of $\phi$ should be the source for the Maxwell equation. Moreover, the terms coupling the time derivatives of $\phi^{(m,n)}$ to the neighboring sectors $\phi^{(m,n+1)}$ and $\phi^{(m,n-1)}$ lead, in every Lorentz frame, to an evolution similar to \eqref{Horig} with photons emitted and absorbed by electrons, as they should be. Obviously, \eqref{LPx} and \eqref{LPy} are Lorentz invariant. The question of consistency of \eqref{LPx}--\eqref{LPg} is subtle and needs further investigation \cite{LT:2021a}. If the system is consistent, then we can say that QED (in this simplified form without positrons) fits very nicely into the framework of the particle-position representation.

Here is what \eqref{LPy} has to do with IBCs. It is well known that part of the Maxwell equations determine the time derivatives while another part is a constraint. In any fixed Lorentz frame, the constraint equations contained in \eqref{Maxwell} read
\be\label{constraint}
\nabla\cdot \vB =0\,,~~~\nabla\cdot \vE=-4\pi J_0
\ee
with $\vB^i=\varepsilon^{ijk}F_{jk}$ and $\vE_i=F_{0i}$. If $J_0(\vy)=e\delta^3(\vy)$, then integration of the second equation over a ball $B_r(\vzero)$ of radius $r$ yields, by the Ostrogradski--Gauss integral theorem,
\be\label{Gauss}
-4\pi e
= \int\limits_{B_r(\vzero)} \! d^3\vy\: (-4\pi e)\delta^3(\vy) 
= \int\limits_{B_r(\vzero)} \! d^3\vy\: \nabla\cdot \vE(\vy) 
= \int\limits_{\SSS^2} d^2\vomega\: r^2\vomega\cdot \vE(r\vomega)\,.
\ee
Suppose that in the limit $r\searrow 0$, the radial component $E_r=\vomega\cdot\vE$ of $\vE$ does not depend on the direction $\vomega$ any more. Then \eqref{Gauss} yields that
\be
\lim_{r\searrow 0} r^2 E_r(r\vomega) = -e.
\ee
For the wave function $\phi$, this implies the following. For simplicity, consider a single $x$-particle fixed at the origin. Then (because $\gamma^0\gamma^0=1$) the second equation of \eqref{constraint} becomes, when all time coordinates are set to 0,
\be\label{ibc2}
\lim_{r\searrow 0} r^2 E_r(\vx=\vzero,\vy_1...\vy_{n-1},r\vomega) = -e\:  \phi^{(1,n-1)}_{\widehat{\mu_n}}(\widehat{y_n})
\ee
with $E_r$ the appropriate $\vy_n$-derivative of $\phi^{(1,n)}$,
\be
E_r(\vx=\vzero,\vy_1...\vy_{n-1},r\vomega) 
= \sum_{i=1}^3 \omega_i \Bigl(\partial^{~}_{y_n,0} \phi^{(1,n)}_{\mu_n=i} - \partial^{~}_{y_n,i} \phi^{(1,n)}_{\mu_n=0}\Bigl)(\vx=\vzero,\vy_1...\vy_{n-1},r\vomega)\,.
\ee
Equation \eqref{ibc2} is an IBC: It is a relation between values (or asymptotics) of $\phi$ (or its derivatives) at $r=0$ and values of $\phi$ at the configuration with one photon removed, i.e., at the configuration related through the absorption of a photon. Note also that the IBC is gauge invariant, as it is expressed in terms of components of $F_{\mu\nu}$, not $A_\mu$. (On earlier occasions \cite{TT15a,LSTT17}, I had written that Landau and Peierls had the first IBC. Thinking again about it, I find that statement too strong, as they wrote their constraint condition with delta distributions, not with boundary conditions.) 

I see two main open problems with the approach of Landau and Peierls (which might be connected): the status of the gauge condition and the Born rule for photons. What is puzzling about the gauge condition \eqref{LPg} is that while Landau and Peierls thought of it as a matter of convenience, it seems that we cannot simply dispense with any gauge condition at all without obtaining too many solutions to the system comprising only \eqref{LPx} and \eqref{LPy} \cite{LT:2021a}. It also seems that the empirical predictions of the theory (even though it is a simplified model that should not be expected to be fully empirically adequate) will depend on the choice of gauge condition, as if there was one ``correct'' condition. But which one would that be, and why? The other problem deserves a section of its own.

\subsection{The Problem of Born's Rule for Photons}
\label{sec:photon}

While it is widely agreed that the quantum state of a single free photon is mathematically equivalent to a (complexified) Maxwell field $F_{\mu\nu}$, there is no agreed-upon answer to what the Born rule for a single photon would say, i.e., for how to compute the probability density $\rho$ in position space from the wave function $F_{\mu\nu}$. For comparison, for a Dirac wave function $\psi:\RRR^4\to\CCC^4$, the probability density $\rho_\mathrm{D}$ or, equivalently, the time component $j^0_\mathrm{D}$ of the probability current 4-vector $j^\mu_\mathrm{D}$, is determined by
\be
j^\mu_\mathrm{D}(x) = \overline{\psi}(x) \, \gamma^\mu \, \psi(x)\,.
\ee
To be sure, there is a convincing way of computing the photon probability current $j^\mu$ provided that the photon wave function (or, equivalently, Maxwell field) $F_{\mu\nu}$ is a plane wave or at least a local plane wave (i.e., a function such that every point has a small neighborhood on which the function is to a good degree of approximation a plane wave): Suppose
\be\label{plane}
A_\mu(x)=a_\mu \, e^{ik_\lambda x^\lambda}\,,~~~
F_{\mu\nu}(x) = 2i \,a_{[\mu}\, k_{\nu]} \, e^{ik_\lambda x^\lambda}
\ee
with future-lightlike $k^\mu$ and $k^\mu a_\mu=0$ to fulfill the free Maxwell equation $\partial^\mu F_{\mu\nu}=0$. If many photons have the same wave function, a classical regime with electromagnetic field $F_{\mu\nu}$ (or something closely related) should apply. Classically, the energy-momentum density is given, up to constant factors, by the tensor field
\be\label{Tdef}
T_{\mu\nu} = \Re\bigl[ F_{\mu\lambda}^* F_\nu^{~\lambda}\bigr] - \tfrac14 g_{\mu\nu} F^*_{\lambda\rho}F^{\lambda\rho} \,,
\ee
which for the plane wave \eqref{plane} amounts to
\be\label{Tplane}
T_{\mu\nu} =a^*_{\lambda}\, a^\lambda \, k_{\mu}k_\nu
\ee
If each photon has momentum $\hbar k^\mu$ according to the de Broglie relation, and if photon number density is proportional to the probability density $j^\nu$ for each photon, then the energy-momentum density should be
\be
T_{\mu\nu} = \hbar k_\mu j_\nu
\ee
up to a constant factor. So, we can obtain $j^\mu$ by comparison with \eqref{Tplane},
\be\label{jplane}
j^\mu = a_\lambda^* a^\lambda\, k^\mu
\ee
up to a constant factor.

However, not every Maxwell field is a local plane wave (even if perhaps in most present-day experiments it is a local plane wave), so the question remains, which law does determine $j^\mu$ in general? Of the desired law, I would expect these properties:
\begin{enumerate}
\item The expression is quadratic in $A_\mu$ and its derivatives.
\item The expression is local, i.e., $j^\nu(x)$ depends only on $A_\mu$ and its derivatives \emph{at $x$}.
\item $j^\mu$ is future timelike-or-lightlike.
\item $\partial_\mu j^\mu=0$ if $A_\mu$ obeys the free Maxwell equations.
\item For a plane wave, $j^\mu$ agrees with \eqref{jplane} up to a constant factor.
\item No choices need to be made, i.e., if the law requires a special gauge or Lorentz frame then it also specifies this gauge or Lorentz frame.
\item The law can be generalized to curved space-time.
\end{enumerate}
If the law involves $A_\mu$, then it might also involve a gauge condition that selects a particular gauge and replaces \eqref{LPg}, thereby also solving the problem mentioned at the end of Section~\ref{sec:LP}.
Several proposals for $j^\mu$ have been made (e.g., \cite{lp:1930,BB94,KTZ18}), but none of them satisfies all of the properties above, so I am hesitant to accept any of them, even though some of them may be useful approximations. It seems to me that the correct answer to the question has not been found yet, although it seems like a very basic question.

\section{Detection Time and Boundary Conditions}
\label{sec:detect}

Another boundary condition on wave functions deserves to be mentioned: an absorbing boundary condition used for formulating Born's rule on a timelike 3-surface $\Sigma$. Consider a 2-surface $\sigma$ in 3-space and suppose detectors are placed along $\sigma$ waiting for a quantum particle to arrive. When and where on $\sigma$ will the particle be detected? In other words, where on the timelike 3-surface $\Sigma = [0,\infty)\times \sigma$ in 4-space-time will it be detected? (Assuming the initial conditions are set up on the surface $\{x^0=0\}$.) The problem of computing the probability distribution of the detection point $X\in\Sigma$ is known as the time-of-arrival problem, although it would perhaps be more accurately called the time-of-detection problem. 

Let me focus again on the non-relativistic case, and let us assume that $\sigma$ is the boundary of a 3-region $\Omega$; for simplicity, I will ignore the possibility that no detector ever clicks. General measurement theory yields that there is a POVM (positive-operator-valued measure) $E(\cdot)$ on $\Sigma$ such that the probability distribution of $X$ is given by
\be
\PPP(X\in B) = \scp{\psi_0}{E(B)|\psi_0}
\ee
for every subset $B$ of $\Sigma$ and every initial wave function $\psi_0$, but it does not tell us in an accessible way which operator $E(B)$ is. 

However, heuristic considerations \cite{detect-rule} suggest the following simple rule for calculating $\PPP(X\in B)$ for idealized detectors, called the \emph{absorbing boundary rule} \cite{Wer87,detect-rule}, and this is what I want to explain in this section. In $L^2(\Omega)$, solve the Schr\"odinger equation
\be\label{Schr}
i\hbar\frac{\partial\psi}{\partial t} = -\tfrac{\hbar^2}{2m}\nabla^2\psi(\vx) + V(\vx) \, \psi(\vx)
\ee
subject to the boundary condition
\be\label{abc}
\vn(\vx)\cdot \nabla\psi(\vx) = i\kappa \psi(\vx)~~~\text{for all }\vx\in\sigma=\partial\Omega
\ee
with $\vn(\vx)$ the unit normal vector to $\sigma$ at $\vx$ and $\kappa$ a given constant such that $\hbar^2\kappa^2/2m$ is the energy of maximal efficiency of the detectors. Then
\be\label{PPPX}
\PPP(X\in B) = \int_B \! dt\, d^2\vx \: \vn(\vx)\cdot \vj(\vx,t)
\ee
with $\vj$ the usual probability current,
\be
\vj= \tfrac{\hbar}{m} \Im [\psi^* \nabla \psi]\,.
\ee

Due to the boundary condition \eqref{abc}, the integrand in \eqref{PPPX} is actually non-negative, as a probability density should be:
\begin{align}
\vn(\vx)\cdot \vj(\vx,t) 
&= \tfrac{\hbar}{m}\Im \bigl[\psi(\vx,t)^*\: \vn(\vx)\cdot \nabla \psi(\vx,t)\bigr] \\[1mm]
&= \tfrac{\hbar}{m}\Im \bigl[\psi(\vx,t)^* \: i\kappa \psi(\vx,t)\bigr] \\[1mm]
&= \tfrac{\kappa\hbar}{m} |\psi(\vx,t)|^2 \geq 0\,.
\end{align}
This means that on the boundary, the current always points outwards, not inwards, so that the particle can cross the boundary only outwards, not inwards. It also implies that the $L^2$ norm of $\psi_t$ is not constant but shrinks with time,
\be
\frac{d}{dt} \|\psi_t\|^2 = \frac{d}{dt} \int_\Omega \! d^3\vx\: |\psi(\vx,t)|^2 \leq 0\,.
\ee
Thus, the time evolution is not unitary (but is what is called a contraction semi-group). It is known \cite{detect-thm} that the time evolution is well defined, i.e., that the Schr\"odinger equation \eqref{Schr} with the boundary condition \eqref{abc} possesses a unique solution for every $\psi_0\in L^2(\Omega)$. This fact reflects the intuitive idea that any part of the wave function $\Psi$ of the object $\vx$ and detectors together that has crossed $\sigma$ in the $\vx$ variable has decohered, i.e., cannot any longer form a coherent superposition with the part that has not crossed $\sigma$ yet. 

That is, the boundary condition here models the effect of the presence of the detector on the wave function in $\Omega$ by precluding backflow of any part of the wave function from outside of $\Omega$. It can be shown \cite{detect-rule,detect-thm} that this rule indeed defines a POVM $E(\cdot)$. It would be of interest to study whether the absorbing boundary rule can be derived from a microscopic quantum-mechanical model of the detectors.

There are several parallels between IBCs and the absorbing boundary condition (ABC) \eqref{abc}. Both are conditions on the wave function. Both are perhaps best understood in terms of their effects on probability transport. Both have been studied only rather recently. And while the IBC provides a way around the UV divergence problem, the ABC provides a way around the quantum Zeno effect that was long through to make impossible any mathematically clean concept of a ``hard'' detector, i.e., one that detects the particle as soon as it reaches the surface $\sigma$.

\paragraph{Conclusions.}
I have given a brief overview of recent research about two kinds of boundary conditions, interior-boundary conditions and absorbing boundary conditions. Both are boundary conditions on wave functions in the particle-position representation, the former for the time evolution with particle creation at point sources, the latter for the time evolution in the presence of hard detectors. As IBCs provide a way out of the UV divergence problem often encountered in connection with particle creation, they are of interest in QFT. They have been proven to work in the non-relativistic case, while the relativistic case remains a field for future research. I have outlined some aspects of how this research might proceed.

\end{document}